\documentclass[aps,twocolumn,preprintnumbers,amsmath,amssymb]{revtex4}

\usepackage{dcolumn}
\usepackage{bm}
\usepackage{graphics,subfigure,xcolor}
\usepackage{braket}
\usepackage{comment}
\usepackage{mathtools}
\usepackage{wrapfig}
\usepackage{microtype}
\usepackage{multirow}
\usepackage{booktabs}

\newcommand{\Tr}{\mathrm{Tr}}
\newcommand{\ad}{a^{\dag}}
\newcommand{\bd}{b^{\dag}}
\usepackage{graphicx}

\begin{document}

\title{Quantum limits to gravity estimation with optomechanics}

\author{F. Armata$^1$}

\author{ L. Latmiral$^{1}$}

\author{A. D. K. Plato$^1$}

\author{M. S.  Kim$^{1,2}$}

\affiliation{$^1$QOLS, Blackett Laboratory, Imperial College London, London SW7 2BW, United Kingdom\\ 
$^2$Korea Institute of Advanced Study, Dongdaemun-gu, Seoul, 02455, South Korea}

\begin{abstract}
We present a table-top quantum estimation protocol to measure the gravitational acceleration $g$ by using an optomechanical cavity. In particular, we exploit the non-linear quantum light-matter interaction between an optical field and a massive mirror acting as mechanical oscillator. The gravitational field influences the system dynamics affecting the phase of the cavity field during the interaction. Reading out such a phase carried by the radiation leaking from the cavity, we provide an estimate of the gravitational acceleration through interference measurements. Contrary to previous studies, having adopted a fully quantum description, we are able to propose a quantum analysis proving the ultimate bound to the estimability of the gravitational acceleration and verifying optimality of homodyne detection. Noticeably, thanks to the light-matter decoupling at the measurement time, no initial cooling of the mechanical oscillator is demanded in principle.
\end{abstract}

\flushbottom
\maketitle
\noindent
\section{Introduction}

Accelerometers are crucially important in a variety of technological applications; consumer electronic products such as cameras and modern smartphones, airbag deployment systems, seismology, as well as aerospace and inertial navigation being just a few of the most widely known examples \cite{krishnan2007}. In recent years, new classes of devices based on quantum principles have been developed to provide clear advantages over conventional systems. One of the most notable examples is that of cold atom gravimeters \cite{muller2008, bidel2013, abend2016}, which promise much improved stability over classical technologies.\\
The key principle consists in the sensitive measurement of the displacement of a test mass, which can be implemented via capacitive \cite{acar2003}, piezo-electric (or resistive) \cite{tadigadapa2009}, optical \cite{krishnamoorthy2008} or atom-based \cite{abend2016} structures. Increasing demand in industrial and consumer applications has recently pushed research towards the miniaturization of accelerometer devices. It is in this direction that microelectromechanical systems (MEMS) accelerometers based on silicon technology have become very attractive. At the same time, the sensitivity of these devices has allowed for a range of applications, e.g. the measurement of the Earth's tides \cite{middlemiss2016}. However, typically such systems are operated in a regime where the test mass displacement is inversely proportional to the square of the mechanical frequency, $\sim a/\omega^2$. This undermines the resulting performance since high-speed operations rely on large resonance frequencies. It is to overcome such difficulties and optimise the tradeoff between noise sensitivity and bandwidth that optomechanical accelerometers, which make use of ultrasensitive displacement readouts using a photonic-crystal nanocavity, have been experimentally implemented \cite{krause2012}.

On the other hand, optomechanical systems have been widely used in a range of different platforms from the quantum-to-classical transition \cite{penrose2003,armata2016} and the test of fundamental physics \cite{bahrami2014} to the preparation of nonclassical states\cite{bose1997,wollman2015}.
Thanks to the very particular light-matter interaction, optomechanical cavities provide an excellent framework to investigate the quantumness of massive objects. The system dynamics can be reconstructed by reading out the light field escaping from the cavity through high precision interference measurements below the standard quantum limit \cite{teufel2009}, i.e. the position uncertainty in the quantum ground state of the mechanical object. A straightforward implementation of this procedure is featured in force sensing \cite{caves1980,braginsky1995,gavartin2012}, which has paved the way to many investigations on the quantum nature of light and macroscopic objects \cite{safavi2013}. Indeed, optomechanical cavities have been recently proposed to witness the observation of exotic dynamical modifications \cite{pikovski2012, latmiral2016} as well as gravitational induced decoherence \cite{penrose1998}. It is in this direction that we propose a fully quantum investigation of the theoretical effectiveness of measuring the gravitational acceleration $g$ using optomechanics.

In particular, we present a method based on the measurement of the phase shift of an optical field after its interaction with a massive quantum oscillator. Relying on the initial preparation of the light in a coherent state, the protocol is resilient to optical losses \cite{aspelmeyer2014}, only requiring a feasible cooling of the oscillator. This is possible thanks to the achievement of a closed loop in the mechanical phase space, which makes the protocol independent of the initial thermal noise. We derive the ultimate quantum bound on the precision that can be reached in probing the gravitational field and show that it is obtainable with the most established measurements on an optical field, i.e. homodyne and heterodyne detection. Rather than providing a feasible experimental constraint on the resolution of gravity estimation protocols, we aim to analyze the quantum limitations arising from fundamental physics, which could provide useful insights on the understanding of the gravitational force. We apply tools from quantum estimation theory and calculate the Quantum Fisher Information (QFI) and the Fisher Information (FI) quantifying the performance of our method. We finally evaluate the corresponding signal-to-noise ratio to further discuss the optimal accuracy of experimental measurements.

\section{\label{sec: Description of the model} Description of the model}
We consider the system depicted in Fig 1 where an optomechanical cavity is vertically oriented
with the two mirrors parallel to the ground. The upper mirror is in a fixed position, while the other, of mass $m$, is free to move
under the effect of a mechanical harmonic potential, of frequency $\omega$, and the gravitational field. We take the length of the cavity at equilibrium to be $L$.  If an optical field in resonance with the cavity frequency $\omega_c$ is injected, the radiation pressure force displaces the mirror by a length $x$ from its equilibrium position (where we assume $x/L \ll 1$). Generalizing the approach in Refs. \cite{Law1995, mancini1997}, the quantized Hamiltonian of the system reads $H=H_{\rm f}+H_{\rm m}+H_{\rm int}+\Delta U$, where $H_{\rm f}= \hbar \omega_c \ad a$, $H_{\rm m}= \frac{p^2}{2m}+\frac{m}{2}\omega^2x^2$ and $H_{\rm int}= -(\hbar\omega_c/L) \ad a x$. Here, $\ad$, $a$ ($\bd$, $b$) are the creation and annihilation operators of the field (mirror), and $x=\sqrt{\hbar/2m\omega}(\bd +b)$ and $p=\sqrt{\hbar m\omega/2}i(\bd-b)$ are the mechanical position and momentum operators. The final term contains the contribution from the gravitational field, and to first order reads $\Delta U = -gmx$, where $g$ is the local gravitational acceleration. Higher order contributions can give additional insight into the density profile of the Earth. As an illustration, we will restrict our analysis to the second order contribution in the displacement (which scales as $x^2/R$) and neglect corrections to the gravity gradient arising because of the influence of the local mass distribution close to the ground. (Deviations from the expected quantities could be evaluated with multi-parameter estimation techniques.) We will thus consider
\begin{equation}
\Delta U\simeq-gmx\left(1+\frac{x}{R}+\mathcal{O}\left(\frac{x^2}{R^2}\right)\right),
\label{Variation-Potential}
\end{equation}
where $R$ is the distance from the lower mirror at equilibrium to the center of the Earth.
In this case, to solve the dynamics, it is convenient to incorporate the quadratic term proportional to $x^2$ into a new unperturbed Hamiltonian $\tilde{H}_{\rm m}$ with frequency $\tilde{\omega}=\sqrt{\omega^2-2g/R}$, so that $H$ can be rewritten as
\begin{equation}\begin{split}\label{H-quantum}
H=&H_{\rm f}+\tilde{H}_{\rm m}-\left(\frac{\hbar\omega_c}{L}+mg\right)x\\
&=\hbar\omega_c \ad a+\hbar\tilde{\omega}\bd b-\hbar(\tilde{g}_0\ad a+S)(\bd+b)\ ,
\end{split}\end{equation}
where $\tilde{g}_0=(\omega_c/L)\sqrt{\hbar/2m\tilde{\omega}}$ is the coupling constant and $S=mg\sqrt{1/(2m\tilde{\omega}\hbar)}$.\\
Equation (\ref{H-quantum}) shows how the gravitational field affects the non-linear interaction by acting directly on the optomechanical coupling between the field and the mechanical oscillator: together with the shift $S$, this provides an enhancement for the final readout of the modified dynamics.\\
Having provided a description of the model, we will now discuss how the gravitational field influences the quantum state of the system.
For convenience, we adopt a frame rotating with the cavity field, where the unitary evolution operator reads \cite{bose1997}
\begin{figure}[h!]
\centering
\includegraphics[width=0.25\textwidth]{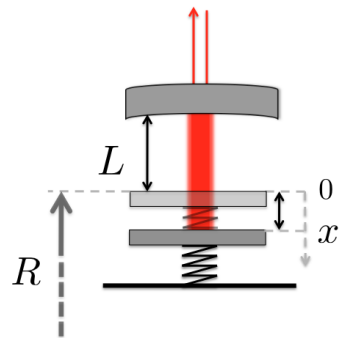}
\caption{Conceptual scheme of the optomechanical cavity under consideration. At the equilibrium, the mechanical mirror is found at a distance $R$ with respect to the center of the Earth (and $L$ with respect to the other fixed mirror) when the cavity is empty. The injected light field displaces it via radiation pressure by an amount $x$.
\label{Model}}
\end{figure}

\begin{equation}\label{evolution-operator}
U(t)=e^{i(\tilde{k}n+\tilde{S})^2(\tilde{\omega}t-\sin\tilde{\omega} t)}e^{(\tilde{k}n+\tilde{S})(\eta\bd-\eta^*b)}e^{-i\bd b\tilde{\omega}t},
\end{equation}
with $n=\ad a$ the number operator of the field, $\eta=(1-e^{-i\tilde{\omega}t})$, $\tilde{k}=\tilde{g}_0/\tilde{\omega}$ and $\tilde{S}=S/\tilde{\omega}$ (see Appendix \ref{Unitary operator}). This generator corresponds to a conditional displacement of the mirror in the mechanical phase space with amplitude $(\tilde{k}\hat{n}_L+\tilde{S})|\eta|$. Let us consider an initial state $|\Psi (0)\rangle=|\alpha,\beta\rangle$, with $|\alpha\rangle_{\rm f}$ and $|\beta\rangle_{\rm m}=|\beta_r+i\beta_i\rangle_{\rm m}$ coherent states for light and mirror respectively, then the state after a time $t$ is given by
\begin{equation}\begin{split}\label{state}
|\Psi(t)\rangle=&\;e^{-\frac{|\alpha|^2}{2}}\sum_n\frac{\alpha^n}{\sqrt{n}!}e^{i(\tilde{k}n+\tilde{S})^2(\tilde{\omega}t-\sin\tilde{\omega}t)}\\
&\times e^{i(\tilde{k}n+\tilde{S})(\beta_r\sin\tilde{\omega}t+\beta_i(1-\cos\tilde{\omega} t))}|n\rangle_{\rm f}\otimes|\gamma (t)\rangle_{\rm m},
\end{split}\end{equation}
where $\gamma (t)=\beta e^{-i\tilde{\omega}t}+(\tilde{k}n+\tilde{S})(1-e^{-i\tilde{\omega}t})$ is the displaced coherent state of the mirror. By computing the expectation values of the position and momentum operators one can verify that the oscillator is driven along a closed circle in phase space,
\begin{equation}\label{circle}\begin{split}
\langle x(t)\rangle&=\sqrt{\frac{2\hbar}{m\omega}}[\beta_r\cos\tilde\omega t +\beta_i\sin\tilde\omega t\\
&\hspace{3cm}+(\tilde{k}N_p+\tilde{S})(1-\cos\tilde\omega t)]\\
\langle p(t)\rangle&=\sqrt{2\hbar m\omega}[\beta_i\cos\tilde\omega t -\beta_r\sin\tilde\omega t\\
&\hspace{3cm}+(\tilde{k}N_p +\tilde{S})\sin\tilde\omega t],
\end{split}\end{equation}
with $N_p=\langle n\rangle$ the intracavity photon number. Equations (\ref{evolution-operator}) and (\ref{state}) tell us that field and mirror are correlated along the evolution (i.e. the system is in an entangled state), and therefore no deterministic measurement can be conducted on a single subsystem without causing the probabilistic state collapse of the other part. On the other hand, the global state becomes separable after every mechanical period, independently of the initial conditions, and acquires an additional geometric-like phase related to the area spanned in phase space. Such independence is a valuable strength of our proposal, which indeed only requires feasible initial cooling. Key to our proposal is the fact that the information on the mechanical dynamics carried by such a phase can be read out by measuring the optical field escaping the cavity. This feature will be discussed in detail in Appendix \ref{Homodyne detection}, where we consider the more general case of an initial thermal state of the mirror.\\

\noindent
\section{Estimation of \MakeLowercase{g}} 
So far, we have shown how the motion of the massive mirror affects the quantum phase that the state of the system acquires along its evolution. We now discuss how to readout the mechanical dynamics through a measurement of the light field. In particular, we will assess the effectiveness of our proposal to estimate the gravitational acceleration $g$ by comparing the precision obtainable with practical measurement schemes with the ultimate bounds set by local quantum estimation theory \cite{paris2009}.

Let us start then from the definition of the QFI relative to the final state (\ref{state}) with respect to the parameter $g$. After a mechanical period $\tau=2\pi/\tilde{\omega}$ the evolution operator in Eq. (\ref{evolution-operator}) acts only on the cavity field, leaving the state of the mirror unchanged. Moreover, if the field is initialized in a pure state $|\alpha\rangle_{\rm f}$ and we evolve the system for a time $\tau$ we obtain $|\psi_g(\tau)\rangle = U(\tau) |\alpha\rangle_{\rm f}$, where $|\psi_g(\tau)\rangle$ is the pure output state of the field and we have made explicit the dependence on $g$. In this particular case the QFI related to a completely positive map acting on the initial state of the light after an interaction lasting a mechanical period reads
\begin{equation}\begin{split}\label{QF}
\mathcal{Q}_g=&4\left(\langle\psi'_{g}|\psi'_{g}\rangle-|\langle\psi'_{g}|\psi_{g}\rangle|^2\right)\\
&=4\big(A^2(4N_p^3+6N_p^2+N_p)\\
&\hspace{2cm}+B^2N_p+2AB(2N_p^2+N_p)\big),
\end{split}\end{equation}
where $|\psi'_{g}\rangle$ is the derivative of the state with respect to the parameter $g$, $A=2\pi \partial \tilde{k}^2/\partial g \simeq -4\pi g_0^2/(\omega^4L_0)$, $B=4 \pi \partial (\tilde{k}\tilde{S})/\partial g \simeq 2\pi (g_0 m/\omega^2)\sqrt{2/\hbar m\omega} \left(1-2g/(\omega^2L_0)\right)$ with $g_0$, $L_0$, respectively, the coupling and cavity length for a horizontally oriented cavity and assuming $L_0\ll R$ (see Appendix \ref{Equilibrium position} for further details). Note, when considering only first order contributions to the gravitational field, then it is clear $A=0$ and the QFI becomes linear in $N_p$. We know from quantum estimation theory that, when the parameter does not influence the measurement aimed at extracting information on the parameter itself \cite{seveso2017}, QFI sets the ultimate lower bound on the estimation precision through the so-called quantum Cram\'er-Rao theorem,
\begin{widetext}
\begin{equation}\begin{split}\label{CRB-t}
{\rm Var}(g)&\geq\frac{1}{\mathcal{M} \mathcal{Q}_g}\gtrsim \frac{L_0^2\omega^{6}}{16\pi^2\mathcal{M}\omega_c^2N_p}\bigg[\left(1-\frac{\hbar\omega_c}{L_0^2m\omega^2}-\frac{2g}{L_0\omega^2}\right)^2\\
&\hspace{5cm}-4N_p\left(\frac{\hbar\omega_c}{L_0^2m\omega^2}\right)\left(1-\frac{3\hbar\omega_c}{2L_0^2m\omega^2}-\frac{2g}{L_0\omega^2}\right)+4N_p^2\left(\frac{\hbar\omega_c}{L_0^2m\omega^2}\right)^2\bigg]^{-1}
\end{split}\end{equation}
\end{widetext}
where $\mathcal{M}$ is the number of measurements performed and the dependence on the mirror mass and frequency, as well as on the number of photons, has been made explicit. In principle, there always exists a POVM whose FI saturates this limit. We will show that feasible and widely adopted measurements schemes, together with adaptive techniques, provide an effective method to probe the gravitational field.
Specifically, we will investigate the performance of homodyne and heterodyne detection at estimating the magnitude of the gravitational force. The former consists of Michelson interferometry where the reference radiation and the signal are derived from the same source. It corresponds to a projection onto the light quadrature operator eigenstates, $X_{\phi}|x\rangle_{\phi}=x|x\rangle_{\phi}$, where $X_{\phi}=x_c \cos\phi+p_c\sin\phi$, and $(x_c, p_c)$ denote, respectively, the {\em position} and {\em momentum} operators for the cavity field, with $\phi$ the phase relative to the local oscillator. More specifically, the light phase shift acquired during the interaction can be analytically from the mean value (after a closed loop in the mechanical phase space)
\begin{equation}\label{phaseg}
\langle a\rangle=\alpha e^{-N_p\left( 1-\cos(4\pi \tilde{k}^2)\right)}e^{i\left[2\pi (2\tilde{k}\tilde{S}+\tilde{k}^2)+N_p\sin(4\pi\tilde{k}^2)\right]}\ .
\end{equation}
\begin{figure}[t!]
\centering
\subfigure[]{
        \label{homodyne_photona}
        \includegraphics[width=0.42\textwidth]{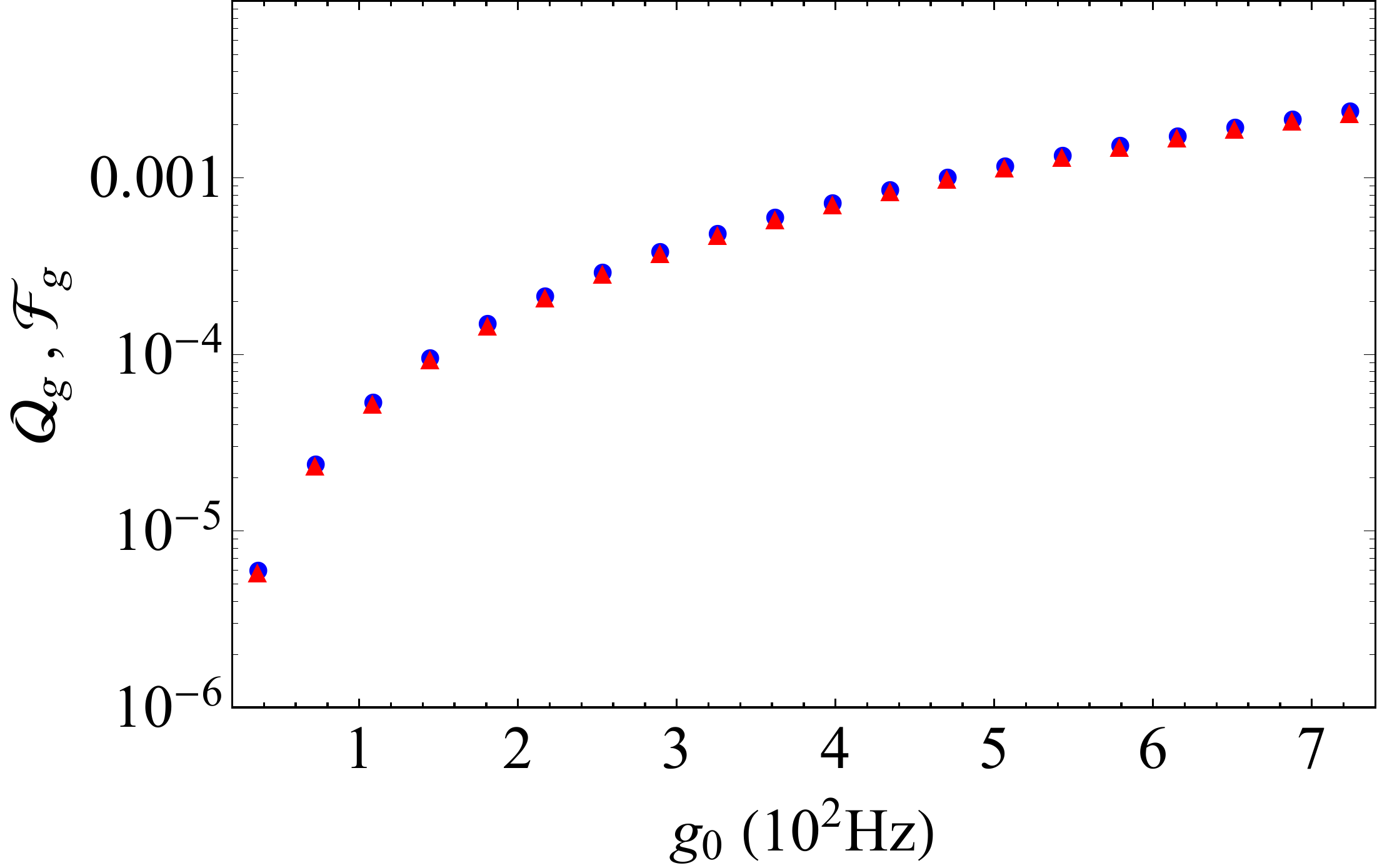} } 
\subfigure[]{
        \label{homodyne_photonb}
        \includegraphics[width=0.42\textwidth]{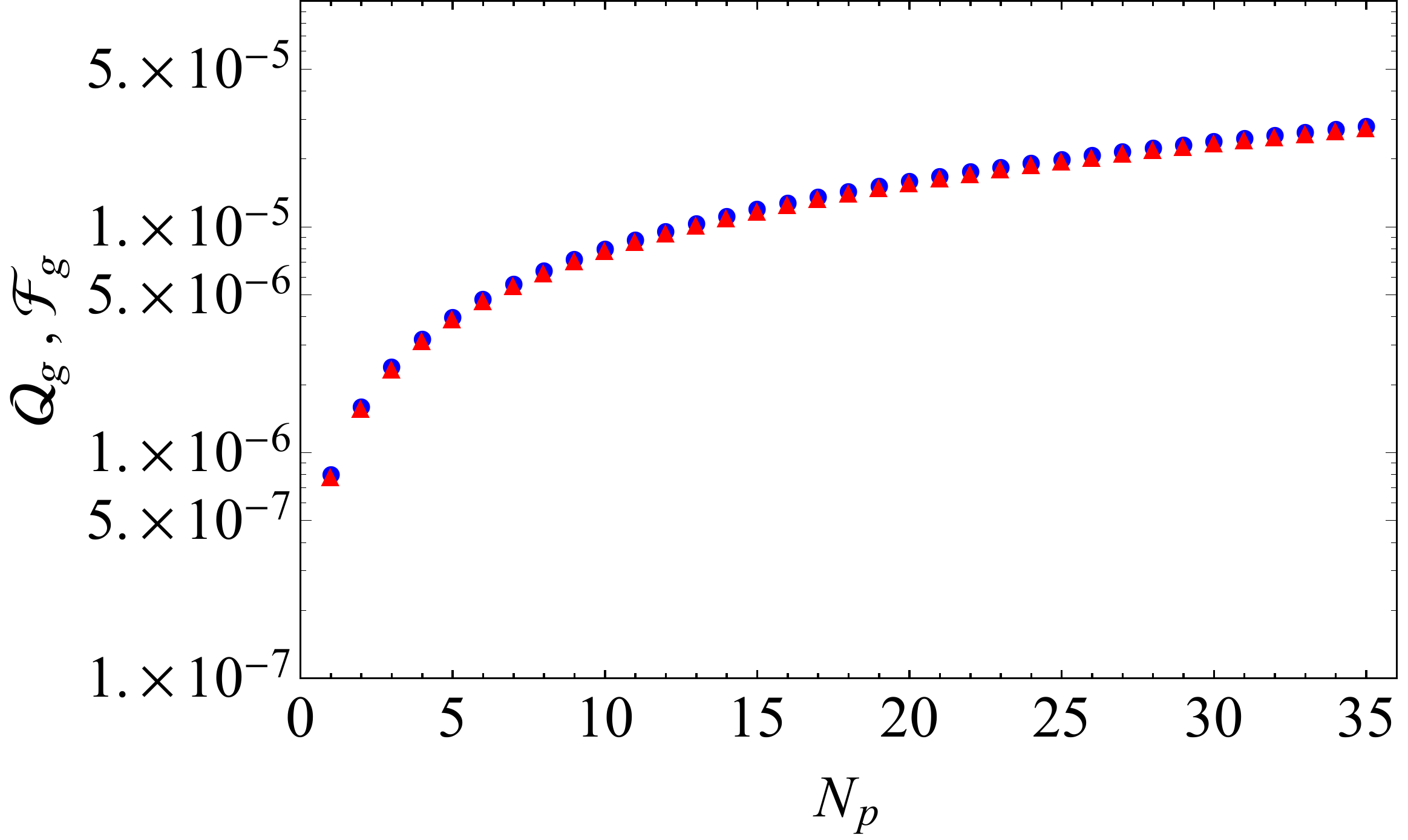} }    
\caption{Quantum Fisher Information (blue dots) and Fisher Information (red triangles) in the case of homodyne detection as functions of the optomechanical coupling $g_0$ (a) and of the average number of photons $N_p$ (b). Cavity parameters are set as  $\omega=2\pi\times 10^7$ Hz, $\omega_c=10^{15}$Hz, $L_0=10^{-4}$m, $m=10^{-7}$Kg. When varying the coupling the number of photons is fixed at $N_p=30$. In both cases the homodyne phase is maximised to $\phi=\pi/2$.}
\label{Fig2}
\end{figure}
The real and imaginary parts of the exponentials in Eq. (\ref{phaseg}), respectively, define the visibility of the interference fringes and the additional phase acquired by the light, the latter being equal to $\varphi=2\pi(\tilde{k}\tilde{S}+\tilde{k}^2)+N_p\sin[4\pi\tilde{k}^2]$. Since $\tilde{k}$ and $\tilde{S}$ are functions of the gravitational field, by inverting the expression of the phase one obtains an indirect measure of $g$. Moreover, with Eq. \eqref{phaseg} at hand, one is able to reconstruct the observable $X_\phi=[ae^{-i\phi}+\ad e^{i\phi}]/\sqrt{2}$.\\
We are now in a position to discuss the estimation precision by comparing the corresponding QFI and FI. It is convenient to first write the quadrature operator eigenstates in the Fock basis as \cite{ferraro2005}
\begin{equation}\label{xphi}
  |x\rangle_{\phi}=e^{-x^2/2}\left(\frac{1}{\pi}\right)^{1/4}\sum_{m=0}^{\infty}\frac{\mathcal{H}_m(x)}{2^{m/2}\sqrt{m!}}e^{-im\phi}|m\rangle,
\end{equation}
where $\mathcal{H}_m(x)$ are the $m$-th Hermite polynomials. Given the expansion of $ |x\rangle_{\phi}$, the FI related to homodyne detection reads
\begin{equation}\begin{split}\label{FI}
\mathcal{F}_g^{\rm hom}&=\int dx\frac{(\partial_{g}p(x|g))^2}{p(x|g)},
\end{split}\end{equation}
where $p(x| g)$ is the conditional probability of obtaining the outcome $x$ in the gravitational field $g$
\begin{equation}\begin{split}\label{pxg}
p(x| g)&=|{}_\phi\langle x|\psi_{g}\rangle|^2\\
&=\frac{e^{-(N_p+x^2)}}{\sqrt{\pi}}\left|\sum_{m=0}^{\infty}\frac{\alpha^m\mathcal{H}_m(x)}{2^{\frac{m}{2}}m!}e^{im\left[\phi-2\pi(\tilde{k}^2m+2\tilde{k}\tilde{S})\right]}\right|^2\ .
\end{split}\end{equation}
In Fig.\ref{Fig2} we show the $\mathrm{QFI}$ and the $\mathrm{FI}$ for homodyne detection as functions of the optomechanical coupling $g_0$ and the number of photons $N_p$.
On the one hand, the precision increases with respect to both the characteristic parameters of optomechanical cavities ($g_0$ and $N_p$), witnessing the effectiveness of our scheme at performing gravity measurements. On the other hand, the ratio $\mathcal{Q}_g/\mathcal{F}_g=1$ demonstrates the optimality of the light interference measurement protocol.

Let us now explore another widely adopted estimation scheme: heterodyne detection. This corresponds to a projection onto a coherent state $|\xi\rangle$ whose conditional probability is given by
\begin{equation}\begin{split}\label{p-eta}
p(\xi|g)=& |\langle \xi | \psi_g\rangle |^2\\
&=e^{-(N_p+|\xi|^2)}\left|\sum_{m=0}^{\infty}\frac{(\alpha \xi^*)^m}{m!}e^{-i2\pi m(\tilde{k}^2m+2\tilde{k}\tilde{S})}\right|^2,
\end{split}\end{equation}
where, contrary to homodyne detection, the dependence on the phase parameter $\phi$ has dropped out.
The FI can be computed by integrating over the whole complex plane spanned by coherent states,
\begin{equation}
\mathcal{F}^{\rm het}_g=\frac{1}{\pi}\int d^2\xi \;\; \frac{(\partial_{g}p(\xi|g))^2}{p(\xi|g)}.
\end{equation}
Our numerical results show that the effectiveness of heterodyne measurement, quantified by the ratio between FI and QFI, does not depend on any parameter, and is found to be $\mathcal{F}^{\rm het}_g/\mathcal{Q}_g \simeq 0.88$. We thus  conclude that it is more convenient to perform a homodyne measurement on the cavity field to estimate gravity with higher precision, and nearly quantum limited.

\section{Discussion}

Finally, we should point out that the FI alone is not sufficient to guarantee the quality of a measurement scheme, particularly in the case of phase measurements, where the power to discriminate between very small phase shifts is usually required. Indeed, in such situations the \emph{signal-to-noise ratio} is an important figure of merit used to assess efficient metrology. It quantifies how the genuine contribution of the desired signal compares to the intrinsic noise typical of the quantum measurement under consideration.
In particular, we can write the signal-to-noise ratio relative to our proposal $R_g$ and derive an upper bound as
\begin{equation}
R_g=\frac{g^2}{{\rm Var}(g)}\leq g^2 \mathcal{M} Q_g \:,
\end{equation}
where $Q_g$ denotes the QFI for the parameter of interest (or the FI in case of non-optimal measurements). An important requirement for efficient metrology is to obtain a large signal-to-noise ration with a reasonable number of experimental runs, which translates into $g^2 M \mathcal{Q}_g \gg 1$, where we have taken the QFI as the upper bound for the inverse of the variance of a single measurement.\\
Substituting common up-to-date values for the cavity parameters, {\em e.g.} $N_p \sim 10^5$, $\omega=2\pi\times 10^5$Hz, $\omega_c=10^{15}$Hz, $L_0=10^{-5}$m, $m=10^{-7}$Kg and considering $\mathcal{M} \sim 10^4$ experimental runs (which still allows us to use optimal asymptotic estimators, such as the Bayesian or the MaxLik estimator), we obtain $R_g\lesssim 8\times10^{18}$ which corresponds to a theoretical bound for the relative error on the measure of the gravitational field as low as $\Delta g/g = 1/\sqrt{R_g} \gtrsim 3.5\times10^{-10}$  or a sensitivity of $0.1\mu \mathrm{Gal}/\sqrt{\mathrm{Hz}}$. In Table \ref{Tab1} we provide a comparison between our predicted theoretical results and achieved (or predicted) results in a selection of other platforms. We infer that our scheme compares favorably to most of the currently available technologies, in particular optomechanics might offer a table-top, stable and robust framework allowing ultra high-speed measurements.
\begin{table}
\begin{tabular}{llc}
\hline
Platform&
\multicolumn{1}{c}{$\Delta g/g$} &
\multicolumn{1}{c}{$\mu\mathrm{Gal}/\sqrt{Hz}$}\\ \hline
Atom Interferometry  \cite{muller2008}& $1.3\times 10^{-9}$$^{\dag}$&8$^{\dag}$ \\
Superconducting gravimetry \cite{instrument}&$10^{-12}$$^{\dag}$& 0.3$^{\dag}$ \\
Falling corner cube \cite{lacoste}&$2\times 10^{-9}$$^{\dag}$&15$^{\dag}$ \\
Atom-Chip Fountain Gravimeter \cite{abend2016}&$1.7\times10^{-7}$$^{\dag}$&\\
&($7.8\times10^{-10}$$^{\star}$)&$(5.3^{\star}$)\\
Optomechanics&$3.5\times10^{-10}$$^{\star}$&0.1$^{\star}$\\
\hline
\end{tabular}
\raggedright
$^{\dag}$Achieved $^{\star}$Predicted
\caption
{Comparison between the relative accuracy $\Delta g/g$ and the sensitivity $(\Delta g/g)/\sqrt{Hz}$ converted in $\mu\mathrm{Gal}/\sqrt{Hz}$ for different kind of experiments.}
\label{Tab1}

\end{table}

\section{\label{sec:Conclusions} Conclusions}

We have presented a protocol to measure the gravitational acceleration by using an optomechanical cavity. Exploiting the light-matter interaction, we have taken advantage of the light intensity enhancement of the quantum phase acquired by the system during the evolution. We have proposed a precise estimation technique by reading out the mechanical dynamics in a gravitational field via interference measurements on the optical field. Moreover, we have also provided a fully quantum description of the model to assess the ultimate quantum bound on the estimation precision and have shown how feasible and standard measurement techniques (e.g. homodyne and heterodyne detection) perform optimally and saturate the Cram\'er-Rao bound. The Berry-like nature of the phase acquired alongside the evolution by the optical field, together with the light-matter decoupling at the measurement time, ensures our scheme to be robust against initial thermal fluctuations, thus not requiring initial cooling of the mechanics. Our procedure has applications in many contexts where parameter estimation is demanded and can also provide a platform for further investigations on gravitation induced decoherence. Given recent technological progress in miniaturization of optomechanical apparatus, we hope that our investigation will pave the way to portable accelerometers with a range of possible applications, which may complement existing devices.\\
Interestingly, a very similar result was independently achieved and recently reported in Ref. \cite{qvarfort2017}.

\section{Acknowledgements}
The authors wish to thank Marco Genoni for useful discussions on the subject of this paper.
This work was supported by the People Programme (Marie Curie Actions) of the European Unions Seventh Framework Programme (FP7/2 007-2013) under REA grant
agreement no 317232, and a Leverhulme Trust Research Grant (Project RPG-2014-055). 
\section*{Author contribution}

F. A. and L. L. performed the calculations and equally contributed to the work. A. D. K. P. and M. S. K. conceived and supervised the research. All authors analysed the results and co-wrote the manuscript.

\appendix
\noindent
\section{\label{Unitary operator} Unitary operator}
Here we provide some details on the derivation of the evolution operator in Eq. (\ref{evolution-operator}). The procedure follows the one outlined in Refs. \cite{mancini1997, bose1997}. We start by considering the evolution operator $U(t)=e^{-i(\ad a)\omega_ct}e^{-i(\bd b)\tilde{\omega}t+i(\tilde{k}\ad a+\tilde{S})(\bd+b)\tilde{\omega}t}$ relative to the Hamiltonian in Eq. (\ref{H-quantum}) and introduce the translation operator $\mathcal{T}=e^{-(\tilde{k}\ad a+\tilde{S})(\bd-b)}$. We know that the identity $\mathcal{T}f(\{Y_i\})\mathcal{T}^{\dag}=f(\{\mathcal{T}Y_i\mathcal{T}^{\dag}\})$ is satisfied for any function $f$, unitary $\mathcal{T}$ and set $\{Y_i\}$. By using the following transformations $\mathcal{T}b\mathcal{T}^{\dag}=b+(\tilde{k}\ad a+\tilde{S})$, $\mathcal{T}\ad a\mathcal{T}^{\dag}=\ad a$, $\mathcal{T}\bd b\mathcal{T}^{\dag}=\bd b+(\tilde{k}\ad a+\tilde{S})(\bd+b)+\tilde{k}^2(\ad a)^2+2\tilde{k}\tilde{S}\ad a+\tilde{S}^2$,
one can find $\mathcal{T}U(t)\mathcal{T}^{\dag}=e^{-i(\ad a)\omega_c t}e^{-i(\bd b)\tilde{\omega}t}e^{i(\tilde{k}\ad a+\tilde{S})^2\tilde{\omega}t}$. Multiplying on the left by $\mathcal{T}^{\dag}$ and on the right by $\mathcal{T}$,
and subsequently swap the last two exponentials, we get
\begin{equation}\begin{split}
U(t)=&\;e^{-i(\ad a)\omega_ct}e^{i(\tilde{k}\ad a+\tilde{S})^2\tilde{\omega}t}e^{(\tilde{k}\ad a+\tilde{S})(\bd-b)}\\
&\times e^{-(\tilde{k}\ad a+\tilde{S})(\bd e^{-i\tilde{\omega}t}-be^{i\tilde{\omega}t})}e^{-i(\bd b)\tilde{\omega}t},
\end{split}\end{equation}
where we have used $e^{-i(\bd b)\tilde{\omega}t}[\ad a(\bd-b)]e^{i(\bd b)\tilde{\omega}t}=\ad a(\bd e^{-i\tilde{\omega}t}-be^{i\tilde{\omega}t})$. Finally, by applying Baker-Campbell-Hausdorff formula we combine the third and fourth exponentials and retrieve the result given in Eq. (\ref{evolution-operator}).\\

\noindent
\section{\label{Homodyne detection} Mechanical thermal state}
Starting from the state in Eq. (\ref{state}), where the mirror is initially in a coherent state $|\beta\rangle$, the total density operator of the system can be written as
\begin{equation}\begin{split}\label{rhototcoh}
\rho(t)=&\;e^{-|\alpha|^2}\sum_{m,n}\frac{\alpha^n\alpha^{*m}}{\sqrt{n!m!}}e^{i(\tilde{k}^2(n^2-m^2)+2\tilde{k}\tilde{S}(n-m))(\tilde{\omega} t-\sin\tilde{\omega} t)}\\
&\times e^{i\tilde{k}(n-m)(\beta_r\sin\tilde{\omega t}+\beta_i(1-\cos\tilde{\omega} t))}\\
&\times |n\rangle_{\rm f}\langle m|\otimes|\gamma_n(t)\rangle_{\rm m}\langle\gamma^*_m(t)|\ .
\end{split}\end{equation}
We know that a thermal state corresponds to a statistical mixture of coherent states defined by $\rho_{\rm m}=(\pi \bar{n})^{-1}\int d^2\beta\; e^{-|\beta|^2/\bar{n}}|\beta\rangle_{\rm m}\langle\beta|$, where $\bar{n}=1/(e^{\hbar\omega/(k_bT)}-1)$ is the average thermal occupation number. By computing such a weighted average with the expression in Eq. (\ref{rhototcoh}) and tracing out over the mechanical degrees of freedom we obtain the field reduced density matrix for an initial mechanical thermal state,
\begin{eqnarray}\label{rhofieldapp}
\rho_{\rm f}(t)&=&e^{-|\alpha|^2}\sum_{m,n}\frac{\alpha^n\alpha^{*m}}{\sqrt{n!m!}}e^{i(\tilde{k}^2(n^2-m^2)+2\tilde{k}\tilde{S}(n-m))(\tilde{\omega} t-\sin\tilde{\omega} t)}\nonumber \\
&&\times e^{-\tilde{k}^2(n-m)^2(1-\cos\tilde{\omega} t)(2\bar{n}+1)}|n\rangle_{\rm f}\langle m|\ .
\end{eqnarray}
We can verify from Eq.\eqref{rhofieldapp} that the reduced density matrix of the field is independent of the initial thermal fluctuations of the mirror.\\
As discussed in the main text, homodyne detection corresponds to projecting on the quadrature operator $X_{\phi}=(1/\sqrt{2})[ae^{-i\phi}+a^{\dag}e^{i\phi}]$, where $a$ is the optical field operator that exits the cavity. The first step to compute the expectation value of $X_{\phi}$ is therefore to evaluate the mean value of the optical field, which results in
\begin{eqnarray}\label{optical-field}
\langle a\rangle&=&\Tr[a\rho_{\rm f}]\nonumber \\
&=&\alpha e^{-\tilde{k}^2(1-\cos\tilde{\omega} t)(2\bar{n}+1)}e^{-N_p\left\lbrace 1-\cos[2\tilde{k}^2(\tilde{\omega}t-\sin\tilde{\omega}t)]\right\rbrace}\nonumber \\
&&\times e^{i(2\tilde{k}\tilde{S}+\tilde{k}^2)(\tilde{\omega}t-\sin\tilde{\omega}t)}e^{iN_p\sin[2\tilde{k}^2(\tilde{\omega}t-\sin\tilde{\omega}t)]}.
\end{eqnarray}
This gives us the phase acquired by light after a time $t$. We remark that the mirror being initially in a thermal state affects only the visibility of the interference fringes reducing the amplitude of the coherent field \cite{armata2016}. However, no visibility reduction occurs at every mechanical period, i.e. when the measurements are performed.

\noindent
\section{\label{Equilibrium position} Equilibrium position}
As we have discussed, it is key for the accuracy of the estimation protocol to start with the mechanical oscillator at equilibrium. Here, we evaluate such an equilibrium length of the cavity $L$ with respect to the horizontally oriented value $L_0$, assuming a homogeneous spherically symmetric Earth. From the equilibrium condition of the mechanical oscillator defined by $\tilde{\omega}^2x=g(x)$, with $g(x)=GM/(R-x)^2$, $G$ the gravitational constant and $M$ the mass of the Earth, we get $\tilde{\omega}^2x\simeq g-(\partial g/\partial x)x=g-(2GM/R^3)x$. This gives us $x\equiv\delta L=L-L_0\simeq (g/\omega^2)(1-2GM/(\omega^2R^3))$ and therefore, $L\simeq L_0+(g/\omega^2)(1-2GM/(\omega^2R^3))$, where we have neglected terms of the order $\mathcal{O}(GM/(\omega^2R^4))$. At the same order of approximation the optomechanical coupling reads $\tilde{g}_0=(\omega_c/L)\sqrt{\hbar/2m\tilde{\omega}}=g_0(1-GM/(R^2L_0\omega^2))$, with $g_0=(\omega_c/L_0)\sqrt{\hbar/2m\omega}$ the optomechanical coupling for a horizontally oriented cavity.\\

\bibstyle{apsrev4-1}
\bibliography{ProbingGravityV2}

\end{document}